# Effect of Bi Substitution on Thermoelectric Properties of SbSe$_2$-based Layered Compounds NdO$_{0.8}$F$_{0.2}$Sb$_{1-x}$Bi$_x$Se$_2$

Yosuke Goto,[1]* Akira Miura,[2] Chikako Moriyoshi,[3] Yoshihiro Kuroiwa,[3] and Yoshikazu Mizuguchi[1]

[1]Department of Physics, Tokyo Metropolitan University, 1-1 Minami-osawa, Hachioji, Tokyo 192-0397, Japan

[2] Faculty of Engineering, Hokkaido University, Kita 13, Nishi 8 Sapporo 060-8628, Japan

[3] Department of Physical Science, Hiroshima University, 1-3-1 Kagamiyama, Higashihiroshima, Hiroshima 739-8526, Japan

Although SbSe$_2$-based layered compounds have been predicted to be high-performance thermoelectric materials and topological materials, most of these compounds obtained experimentally have been insulators so far. Here, we present the effect of Bi substitution on the thermoelectric properties of SbSe$_2$-based layered compounds NdO$_{0.8}$F$_{0.2}$Sb$_{1-x}$Bi$_x$Se$_2$ ($x \leq 0.4$). The room temperature electrical resistivity is decreased to $8.0 \times 10^{-5}$ $\Omega$m for $x = 0.4$. The electrical power factor is calculated to be $1.4 \times 10^{-4}$ Wm$^{-1}$K$^{-2}$ at 660 K, which is in reasonable agreement with combined Jonker and Ioffe analysis. The room-temperature lattice thermal conductivity of less than 1 Wm$^{-1}$K$^{-1}$ is almost independent of $x$, in contrast to the point-defect scattering model for conventional alloys. The present work provides an avenue for exploring SbSe$_2$-based insulating and BiSe$_2$-based conducting systems.

1. Introduction

The development of highly efficient thermoelectric materials that convert a temperature difference into electricity has been extensively studied for energy harvesting from waste heat.[1–5] In particular, recent advances in the first-principles approach have enabled reliable predictions of electronic structures and transport properties under some assumptions, such as a constant relaxation time.[6–12] Several candidate materials have been theoretically predicted to be efficient thermoelectric materials from a vast number of materials (known and hypothetical), and the performances of some of these predicted compounds have been experimentally demonstrated so far.

Recently, Ochi *et al.* have presented strategies for improving the thermoelectric properties of layered pnictogen chalcogenides REOPnCh$_2$ (RE = rare earth, Pn = pnictogen, Ch = chalcogen) using first-principles calculation.[13] The crystal structures of these compounds are characterized by alternate stacks of PnCh$_2$ conducting layers and REO spacer layers, as schematically shown in Fig. 1. This family of compounds has been mainly studied as a superconductor when Pn = Bi.[14–16], and recently as a topological material,[17] a water-splitting photocatalyst,[18] and a thermoelectric material.[19–23] As an example, a thermoelectric dimensionless figure of merit ($ZT$) of 0.36 at 650 K was obtained for LaOBiSSe.[19] Although $ZT$ for this family of compounds is still moderate, their intrinsically low lattice thermal conductivity, 1–2 Wm$^{-1}$K$^{-1}$ at 300 K[19–23], is attractive for efficient thermoelectric conversion. In particular, a decrease in lattice thermal conductivity due to the rattling motion of Bi has recently been reported, even though these compounds have no oversized cage.[24] Ochi et al. presented guiding principles for improving the power factor of REOPnCh$_2$-type compounds: (1) small spin-orbit coupling, (2) small Pn−Pn and large Pn−Ch hopping amplitude, and (3) small onsite energy difference between the Pn−$p_{xy}$ and Ch−$p_{xy}$ orbitals.[13] Consequently, replacements of Bi and S ions with lighter Sb and heavier Se, respectively, should result in an increase in thermoelectric performance.

We have recently reported the synthesis and thermoelectric properties of layered antimony oxyselenides REOSbSe$_2$ for RE = La and Ce.[25] The crystal structure belongs to the tetragonal $P4/nmm$ space group, as for most REOBiCh$_2$. However, carrier doping to decrease electrical resistivity was still difficult,[26–28] in sharp contrast to BiCh$_2$-based superconducting systems. Here, we demonstrate the synthesis and thermoelectric properties of NdO$_{0.8}$F$_{0.2}$Sb$_{1-x}$Bi$_x$Se$_2$ (RE = Nd). We found that F-doped NdO$_{0.8}$F$_{0.2}$SbSe$_2$ also crystallizes in the tetragonal $P4/nmm$ space group at room temperature, whereas the F-free compound NdOSbSe$_2$ was not obtained. The electrical resistivity is decreased to $8.0 \times 10^{-5}$ Ωm at 300 K as a result of Bi substitution. The room-temperature lattice thermal conductivity of less than 1 Wm$^{-1}$K$^{-1}$ is almost independent of $x$, in contrast to the point-defect scattering model for conventional alloys.

## 2. Experimental Details

Polycrystalline samples of NdO$_{0.8}$F$_{0.2}$Sb$_{1-x}$Bi$_x$Se$_2$ ($x$ = 0, 0.1, 0.2, 0.3, 0.4) were prepared by solid-state reactions. NdO$_{0.8}$F$_{0.2}$SbSe$_2$ was synthesized using dehydrated Nd$_2$O$_3$, a mixture of compounds mainly composed of NdSe and NdSe$_2$ (NdSe–NdSe$_2$ powder), Sb (99.9%), and Se (99.999%) as starting materials. The dehydrated Nd$_2$O$_3$ was prepared by heating commercial Nd$_2$O$_3$ powder (99.9%) at 600 °C for 10 h in air. To obtain the NdSe–NdSe$_2$ powder, Nd (99.9%) and Se in a molar ratio of 2:3 were heated at 500 °C for 10 h in an evacuated silica tube. Because the Nd powder is reactive in air and a moist atmosphere, this process was carried out in an Ar-filled glovebox. Then, a stoichiometric mixture of these starting materials was pressed into a pellet and heated at 700 °C for 15 h in an evacuated silica tube. To obtain dense samples, the powder samples were hot-pressed using a graphite die at 700 °C at 50 MPa for 30 min (S. S. Alloy, PLASMAN CSP-KIT-02121). The hot-pressed samples had a relative density greater than 96%. To obtain the Bi-doped sample, Bi (99.999%) powder was added in the starting materials.

The phase purity and crystal structure of the samples were examined by synchrotron powder X-ray diffraction (SPXRD) performed at the BL02B2 beamline of SPring-8 (proposal number 2018A0074). The diffraction data were collected using a high-resolution one-dimensional semiconductor detector (multiple MYTHEN system).[29] The wavelength of the radiation beam was determined to be 0.495274(1) Å using a $CeO_2$ standard. The crystal structure parameters were refined by the Rietveld method using RIETAN-FP software.[30] The crystal structure was visualized using VESTA software.[31]

The chemical compositions of the obtained samples were examined using a scanning electron microscope (SEM; Hitachi, TM3030) equipped with an energy dispersive X-ray spectrometer (EDX; Oxford, SwiftED3000). The electrical resistivity $\rho$ and Seebeck coefficient $S$ were simultaneously measured using the conventional four-probe geometry (Advance Riko, ZEM-3) in a He atmosphere. Typically, samples were $2 \times 3 \times 9$ mm$^3$ in size and rectangular parallelepiped in shape. The thermal conductivity was calculated using $\kappa = DC_p d_s$, where $D$, $C_p$, and $d_s$ are the thermal diffusivity, specific heat, and sample density, respectively. The thermal diffusivity was measured by a laser flash method (Advance Riko, TC1200-RH). The samples used for the measurements were disks 10 mm in diameter and 2 mm in thickness. The specific heat was measured by a comparison method using differential scanning calorimetry (DSC; Advance Riko, DSC-R) under an Ar atmosphere.

## 3. Results and Discussion

### 3.1 Structural characterization

Figure 2 shows the SPXRD pattern and Rietveld fitting results for $NdO_{0.8}F_{0.2}SbSe_2$ as a representative sample. Almost all the diffraction peaks can be assigned to those of the tetragonal *P*4/*nmm* space group, except for several tiny peaks attributable to NdOF (1.9 wt%) and $Sb_2Se_3$ (2.0 wt%). The lattice parameters are calculated to be $a = 4.04769(1)$ Å and $c = 14.08331(9)$ Å. These are distinctly smaller than those of REOSbSe$_2$ for RE = La and Ce,[25]) because of (1) the partial substitution of smaller $F^-$ to the site of $O^{2-}$ ions (for example, Shannon's six-coordinate ionic radii are $r_{F^-}$ = 133 pm and $r_{O^{2-}}$ = 140 pm), and (2) the smaller ionic radius of $Nd^{3+}$ than those of $La^{3+}$ and $Ce^{3+}$ ($r_{Nd^{3+}}$ = 110.9 pm, $r_{La^{3+}}$ = 116.0 pm, and $r_{Ce^{3+}}$ = 114.3 pm), namely, the lanthanide contraction rule.[32]) Selected bond distances and angles are shown in Fig. 1(b). In particular, the bond distance between Sb and Se is considered to be essential to determine the electrical transport properties because the conduction band is mainly composed of the hybridization between p orbitals of these ions.[13,33]) The Sb ions are in a square pyramidal edge-shared environment of four in-plane selenium ions (Se1) and one out-of-plane selenium ion (Se2). The bond distances are evaluated to be 2.87079(11) Å for Sb-Se1 and 2.5472(13) Å for Sb-Se2 bonds. These are in reasonable agreement with the bond distance in binary $Sb_2Se_3$ containing a distorted square pyramidal configuration, 2.78–3.01 Å for Sb-Se1 and 2.56–2.59 Å for Sb-Se2.[34–41])

Diffraction peaks for Bi-doped samples, $NdO_{0.8}F_{0.2}Sb_{1-x}Bi_xSe_2$ ($x \leq 0.4$), can also be assigned to those of the tetragonal $P4/nmm$ space group, as shown in Fig. S1.[42]) However, the amount of Bi-containing impurity phase increased with $x$, leading to 5.6 wt% of $Bi_2Se_3$ and 2.4 wt% of $Bi_2O_{4-\delta}$ for $x = 0.4$. As shown in Fig. 3, the lattice parameters $a$ and $c$ increased almost linearly with increasing $x$ owing to the larger ionic radius of $Bi^{3+}$ ions (96 pm) than that of $Sb^{3+}$ ions (80 pm). Figure 4 shows the chemical composition ratios of Nd, Sb, Bi, and Se determined using EDX. The results indicate that the chemical compositions of the obtained samples are in reasonable agreement with the nominal stoichiometries of the starting materials. These characterizations of the crystal structure and chemical composition confirm that Bi ions were systematically incorporated in $NdO_{0.8}F_{0.2}Sb_{1-x}Bi_xSe_2$.

*3.2 Thermoelectric properties*

Figure 5(a) shows the temperature dependence of the electrical resistivity. For $x = 0$, the room-temperature electrical resistivity is $3.1 \times 10^{-3}$ Ωm, and it decreases with increasing temperature. This is distinctly lower than that of RE(O,F)SbSe$_2$ for RE = La or Ce. For example, the room-temperature electrical resistivities are $\sim 10^3$ Ωm for LaOSbSe$_2$ and $\sim 1$ Ωm for LaO$_{0.9}$F$_{0.1}$SbSe$_2$. The decrease in resistivity is most likely due to (1) electron doping by $F^-$ substitution to the site of $O^{2-}$ ions, and (2) increased overlapping between Sb p and Se p orbitals via the decreased lattice parameters and bond distance, as described above.

The electrical resistivity was decreased by Bi substitution, leading to $8.0 \times 10^{-5}$ $\Omega$m for $x = 0.4$ at 300 K. Note that semiconducting behavior, namely, a negative temperature coefficient, was still observed for all the samples. At the same time, the Seebeck coefficient was also decreased, as shown in Fig. 5(b), most likely due to an increase in carrier density by the Bi substitution. As depicted in Fig. 5(c), the calculated electrical power factor (PF, $\rho^{-1}S^2$) was found to be $1.4 \times 10^{-4}$ Wm$^{-1}$K$^{-2}$ at 660 K for $x = 0.3$.

Generally, thermoelectric carrier transport is discussed on the basis of the Hall carrier density as well as the electrical resistivity and Seebeck coefficient. However, it is difficult to measure the Hall coefficient of the present sample. The room-temperature Hall coefficient was estimated to be less than $1 \times 10^{-8}$ m$^3$C$^{-1}$. Here, we discuss the carrier transport using combined Jonker and Ioffe analysis.[43–45] On the basis of a parabolic band model, the Seebeck coefficient of a nondegenerate n-type semiconductor is expressed as

$$S = \frac{k}{e}(\ln \sigma - \ln \sigma_0),$$

$$\sigma_0 = N_c e \mu \exp(A),$$

where $k$ is Boltzmann's constant, $e$ is electronic charge, $\sigma$ is electrical conductivity, $N_c$ is the conduction band density of states, $\mu$ is carrier mobility, and $A$ is the transport constant. The maximum value of PF (PF$_{max}$) was evaluated to be

$$\ln PF_{max} \approx -19.33 + \ln \sigma_0.$$

Figure 5(d) shows the Seebeck coefficient versus the logarithm of electrical conductivity. One may expect that the structure of the conduction band is markedly changed owing to Bi substitution, because the conduction band is mainly composed of the hybridization between Sb/Bi p and Se p orbitals. However, the observed data is in reasonable agreement with the slope of $ke^{-1} = 86.15$ $\mu$VK$^{-1}$, which assumes a parabolic band model. ln$\sigma_0$ is estimated to be $10.6 \pm 0.3$ Sm$^{-1}$, and PF$_{max}$ was calculated to be $(1.7 \pm 0.5) \times 10^{-4}$ Wm$^{-1}$K$^{-2}$. The experimentally obtained PF ($1.4 \times 10^{-4}$ Wm$^{-1}$K$^{-2}$ at 660 K) was in reasonable agreement with this assessment.

Figure 6(a) shows the total thermal conductivity as a function of temperature. The total thermal conductivity is generally expressed as a combination of lattice and electronic contributions ($\kappa = \kappa_{lat} + \kappa_{el}$). The Wiedemann–Franz relation, $\kappa_{el} = LT\rho^{-1}$, is used to evaluate the electronic contribution to the thermal conductivity. Temperature-dependent Lorenz numbers ($L$) were calculated within a parabolic band approximation assuming electronic transport dominated by acoustic phonon scattering.[46,47] The room-temperature lattice thermal conductivity was < 1 Wm$^{-1}$K$^{-1}$ for the present samples, as shown in Fig. 6(b). This low lattice thermal conductivity is comparable to those of related compounds, including SbSe$_2$-based or BiCh$_2$-based systems.[19–25] Notably, the room-temperature lattice thermal conductivity is almost independent of the Bi content $x$. This is in contrast to the conventional point defect scattering model for alloys, such as Si$_{1-x}$Ge$_x$, where phonon scattering is described using the mass and strain contrast of an alloy.[48]

The calculated figure of merit $ZT$ is of around 0.06 at 660 K, as shown in Fig. 7. Although the lattice thermal conductivity of less than 1 Wm$^{-1}$K$^{-1}$ for the present samples should be promising for their use as thermoelectric materials, $ZT$ is very low because of their poor power factor. To compare the experimental results with theoretical ones, measurements of the Hall coefficient will be helpful. However, the Hall coefficients of the present samples are very low, as described above. High-magnetic-field instruments will be required to measure the Hall coefficient correctly.

## 4. Summary

We demonstrate the effect of Bi substitution on the crystal structure and thermoelectric properties of the layered antimony oxyselenide NdO$_{0.8}$F$_{0.2}$Sb$_{1-x}$Bi$_x$Se$_2$ ($x \leq 0.4$). Systematic changes in the lattice parameters determined using SPXRD and the chemical composition determined using EDX indicate that the Bi ions were incorporated in NdO$_{0.8}$F$_{0.2}$Sb$_{1-x}$Bi$_x$Se$_2$ solid solution. Both the electrical resistivity and Seebeck coefficient decrease with increasing Bi content $x$. The room-temperature electrical resistivity is decreased to $8.0 \times 10^{-5}$ Ωm for $x = 0.4$. The obtained PF was $1.4 \times 10^{-4}$ Wm$^{-1}$K$^{-2}$ at 660 K, which is in reasonable agreement with PF$_{max}$ evaluated using combined Jonker and Ioffe analysis. The room-temperature lattice thermal conductivity, < 1 Wm$^{-1}$K$^{-1}$, is almost independent of the Bi content. In this work, we provide an avenue for exploring SbSe$_2$-based insulating and BiSe$_2$-based conducting systems.

**Acknowledgments**

We thank Prof. K. Kuroki and Drs. M. Ochi and N. Hirayama of Osaka University for fruitful discussions. This work was supported by JST-CREST (No. JPMJCR16Q6), Grants-in-Aid for Scientific Research (Nos. 15H05886 and 16H04493), and Iketani Science and Technology Foundation (No. 0301042-A), Japan.

*E-mail: y_goto@tmu.ac.jp

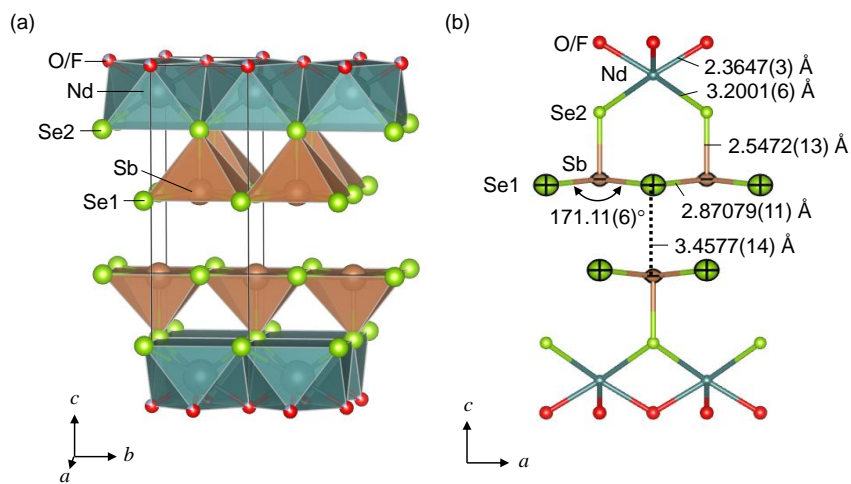

**Fig. 1.** (Color online) (a) Crystallographic structure of NdO$_{0.8}$F$_{0.2}$SbSe$_2$ (tetragonal *P*4/*nmm* space group). The black line denotes the unit cell. Se ions have two crystallographic sites: in-plane (Se1) and out-of-plane (Se2). (b) Selected bond distances and angles for NdO$_{0.8}$F$_{0.2}$SbSe$_2$. For Sb and Se1 atoms, thermal ellipsoids at 90% probability level are also shown.

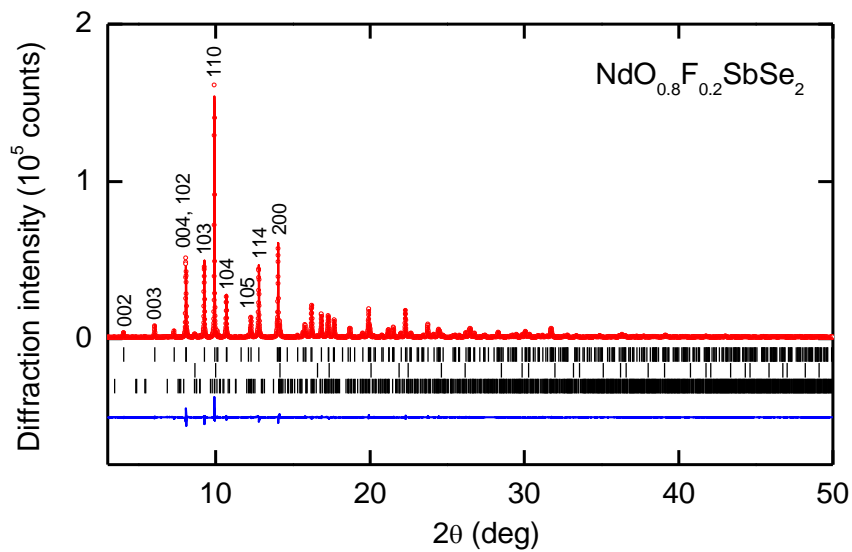

**Fig. 2.** (Color online) Synchrotron powder X-ray diffraction (SPXRD) pattern and results of Rietveld refinement for $NdO_{0.8}F_{0.2}SbSe_2$. The wavelength of the radiation beam was determined to be 0.495274(1) Å. The circles and solid curve represent the observed and calculated patterns, respectively, and the difference between the two is shown at the bottom. The vertical marks indicate the Bragg diffraction positions for $NdO_{0.8}F_{0.2}SbSe_2$ (top), NdOF (middle), and $Sb_2Se_3$ (bottom). Analogous data for Bi-doped samples are shown in Fig. S1 in Supporting Information.

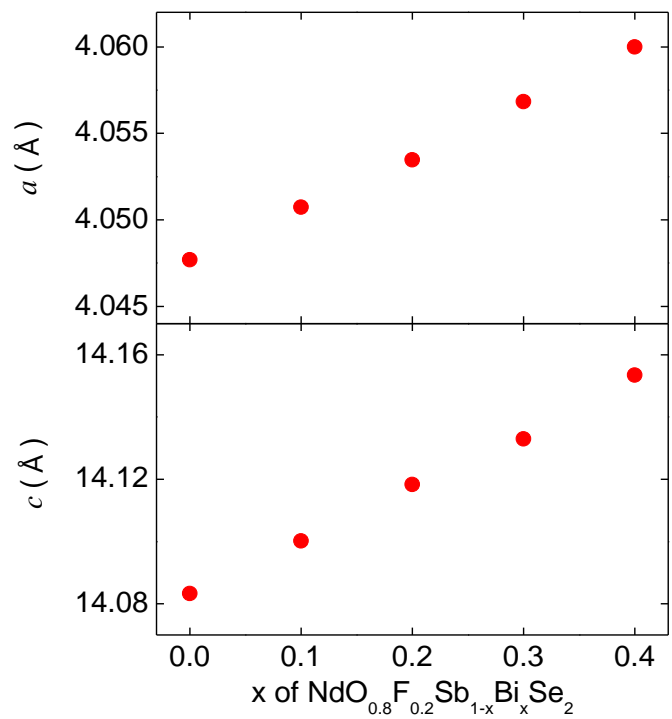

**Fig. 3.** (Color online) Lattice parameters ($a$ and $c$) of $NdO_{0.8}F_{0.2}Sb_{1-x}Bi_xSe_2$.

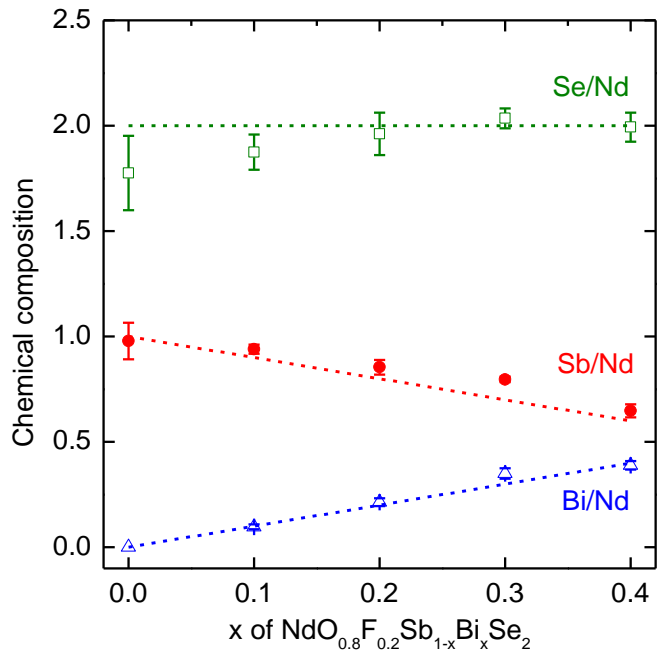

**Fig. 4.** (Color online) Chemical composition determined using EDX for NdO$_{0.8}$F$_{0.2}$Sb$_{1-x}$Bi$_x$Se$_2$. The dashed lines denote the nominal compositions of the starting materials.

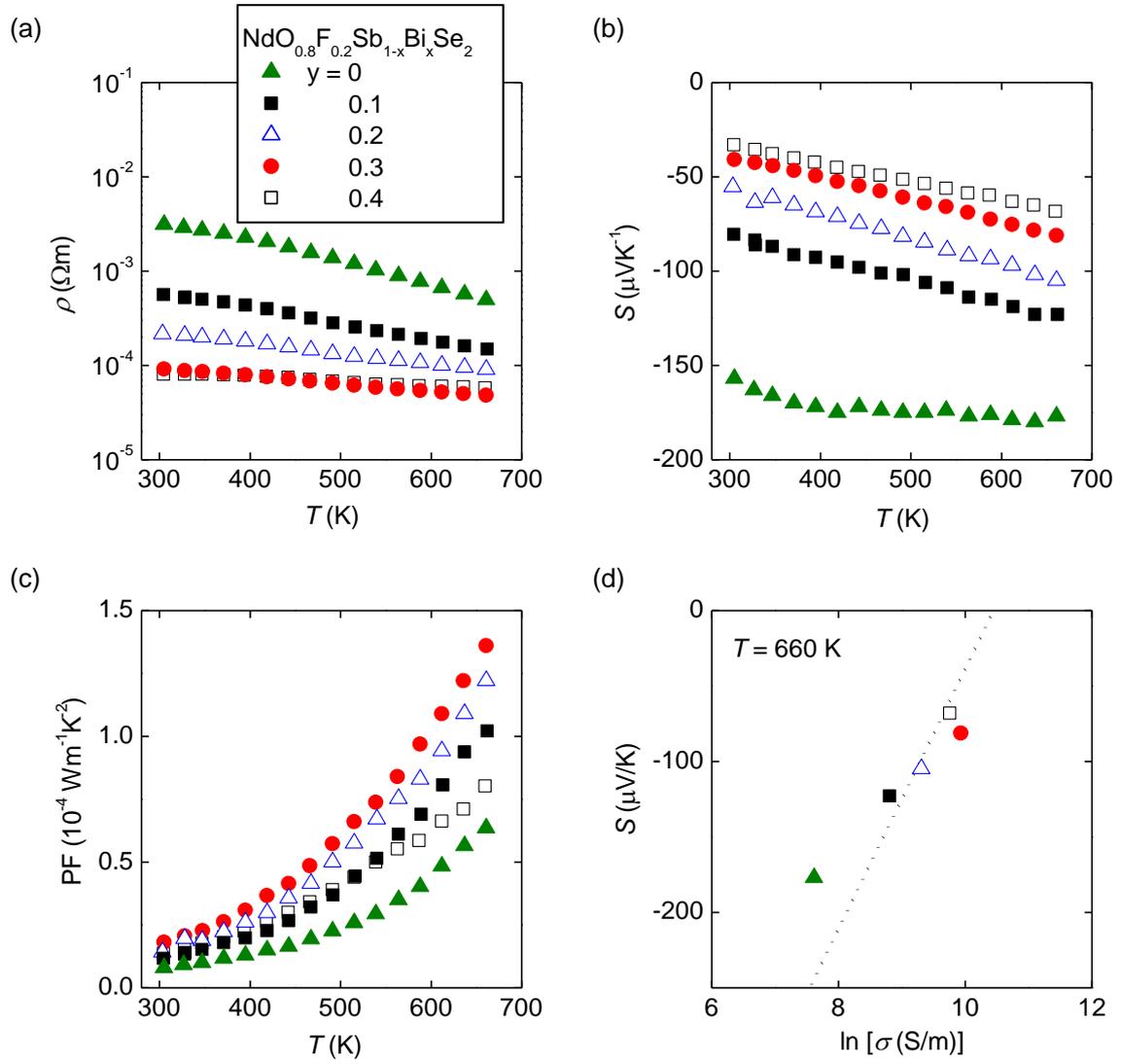

**Fig. 5.** (Color online) Electrical transport properties of NdO$_{0.8}$F$_{0.2}$Sb$_{1-x}$Bi$_x$Se$_2$. (a) Electrical resistivity ($\rho$) versus temperature ($T$), (b) Seebeck coefficient ($S$) versus $T$, (c) power factor (PF) versus $T$, and (d) $S$ versus logarithm of electrical conductivity ($\sigma$), a Jonker plot. The dashed line in panel (d) represents the slope of $ke^{-1} = 86.15$ μVK$^{-1}$.

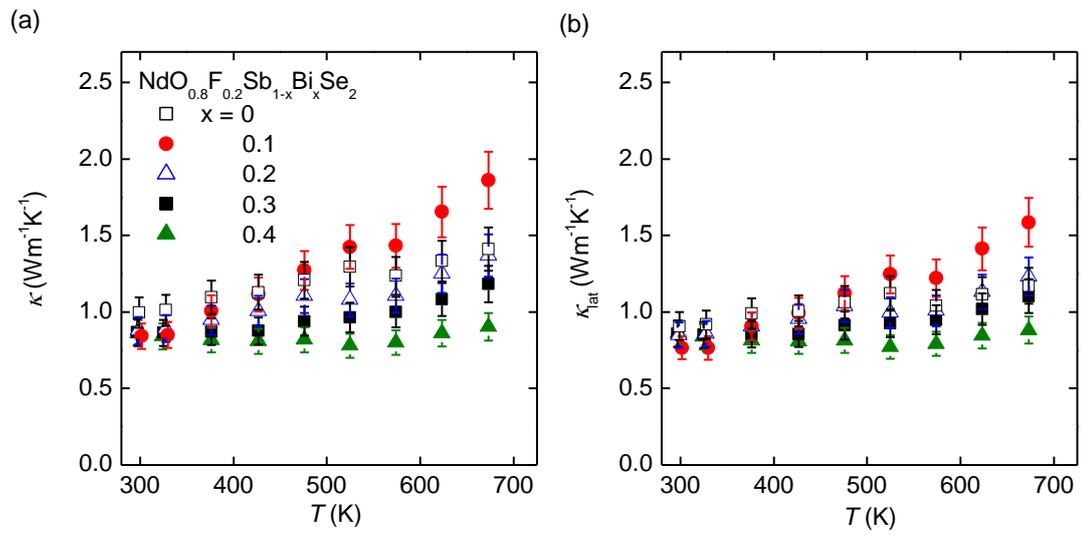

**Figure 6.** (Color online) Temperature ($T$) dependences of (a) total thermal conductivity ($\kappa$) and (b) lattice thermal conductivity ($\kappa_{lat}$) for NdO$_{0.8}$F$_{0.2}$Sb$_{1-x}$Bi$_x$Se$_2$.

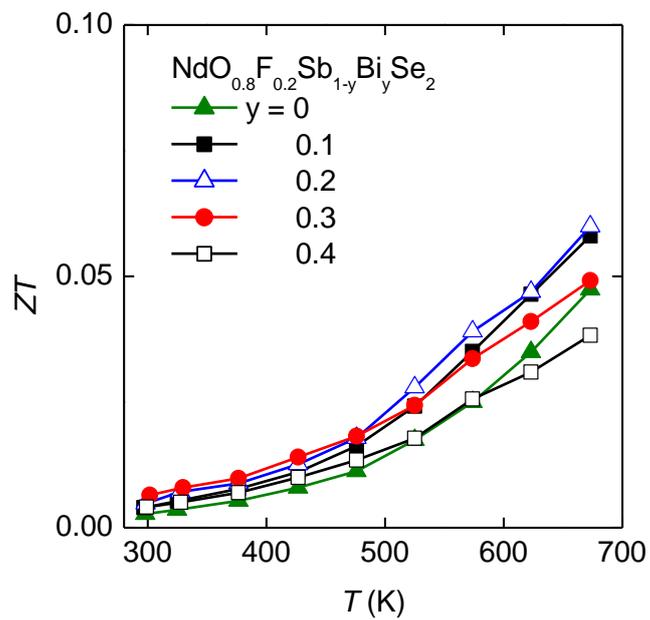

**Figure 7.** (Color online) Dimensionless figure of merit (*ZT*) as a function of temperature (*T*) for NdO$_{0.8}$F$_{0.2}$Sb$_{1-x}$Bi$_x$Se$_2$.

**Table 1.** Crystal structure parameters and reliability factors of $NdO_{0.8}F_{0.2}SbSe_2$ obtained by Rietveld analysis.[a]

| Lattice system | | | | | Tetragonal | |
|---|---|---|---|---|---|---|
| Space group | | | | | $P4/nmm$ (No. 129) | |
| Lattice parameters | | | | | $a = 4.04769(1)$ Å, $c = 14.08331(9)$ Å | |
| atom | site | $g$ | $x$ | $y$ | $z$ | $U$ (Å$^2$) |
| Nd | 2c | 1.0 | 0 | 1/2 | 0.08684(4) | $U_{iso} = 0.00813(15)$ |
| O | 2a | 0.8 | 0 | 0 | 0 | $U_{iso} = 0.0118(18)$ |
| F | 2a | 0.2 | 0 | 0 | 0 | $U_{iso} = 0.0118(18)$ |
| Sb | 2c | 1.0 | 0 | 1/2 | 0.63066(5) | $U_{11} = 0.0244(4)$ |
| | | | | | | $U_{33} = 0.0098(5)$ |
| | | | | | | $U_{eq} = 0.020$ |
| Se(1) | 2c | 1.0 | 0 | 1/2 | 0.38514(8) | $U_{11} = 0.0309(6)$ |
| | | | | | | $U_{33} = 0.0230(9)$ |
| | | | | | | $U_{eq} = 0.028$ |
| Se(2) | 2c | 1.0 | 0 | 1/2 | 0.81153(7) | $U_{iso} = 0.0096(3)$ |
| $R_{wp}$ | | | | | 5.69% | |
| $R_B$ | | | | | 1.15% | |
| GOF | | | | | 2.40 | |

[a]The values in parentheses are standard deviations in the last digits. For Sb and Se(1) sites, anisotropic displacement parameters are refined.

Supporting Information for
# "Effect of Bi Substitution on Thermoelectric Properties of SbSe$_2$-based Layered Compounds NdO$_{0.8}$F$_{0.2}$Sb$_{1-x}$Bi$_x$Se$_2$"

Yosuke Goto,[1]* Akira Miura,[2] Chikako Moriyoshi,[3] Yoshihiro Kuroiwa,[3] and Yoshikazu Mizuguchi[1]

[1]Department of Physics, Tokyo Metropolitan University, 1-1 Minami-osawa, Hachioji, Tokyo 192-0397, Japan

[2] Faculty of Engineering, Hokkaido University, Kita 13, Nishi 8 Sapporo 060-8628, Japan

[3] Department of Physical Science, Hiroshima University, 1-3-1 Kagamiyama, Higashihiroshima, Hiroshima 739-8526, Japan

*E-mail: y_goto@tmu.ac.jp

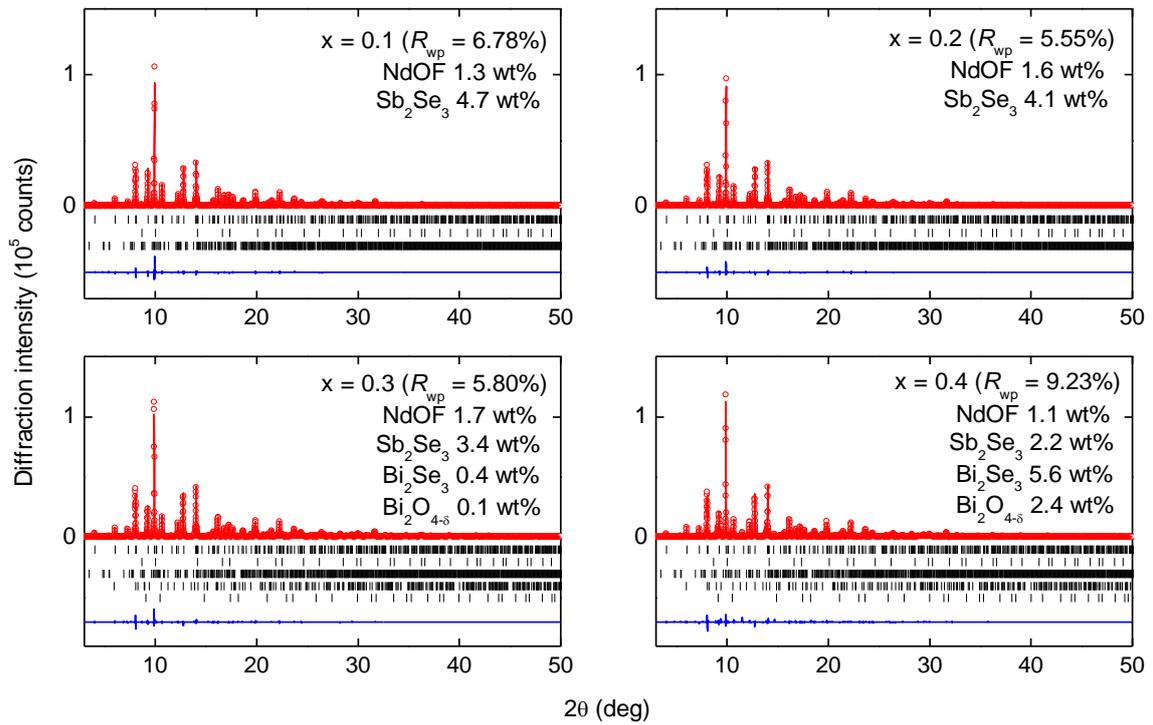

**Fig. S1.** Synchrotron powder X-ray diffraction (SPXRD) pattern and the results of Rietveld refinement for $NdO_{0.8}F_{0.2}Sb_{1-x}Bi_xSe_2$. The circles and solid curve represent the observed and calculated patterns, respectively, and the difference between the two is shown at the bottom. The vertical marks indicate the Bragg diffraction positions for $NdO_{0.8}F_{0.2}Sb_{1-x}Bi_xSe_2$ and impurity phases, NdOF, $Sb_2Se_3$, $Bi_2Se_3$, and $Bi_2O_{4-\delta}$. Amount of impurities is shown in the inset.